\RequirePackage{fix-cm}
\documentclass[twocolumn,epjc3]{svjour3}  %
\smartqed
\RequirePackage{graphicx}
\RequirePackage{mathptmx}
\RequirePackage{amsmath}
\RequirePackage{amsfonts}
\RequirePackage{amssymb}
\RequirePackage{latexsym}
\RequirePackage{widetext}

\newcommand{\be}{\begin{equation}}
\newcommand{\ee}{\end{equation}}
\newcommand{\bea}{\begin{eqnarray}}
\newcommand{\eea}{\end{eqnarray}}

\renewcommand{\baselinestretch}{1.02}

\journalname{Eur. Phys. J. C}
\begin{document}

\title{Inquiring electromagnetic quantum fluctuations about the orientability of space}

\author{N. A. Lemos\thanksref{e1,addr1}
        \and
       \  M. J. Rebou\c{c}as\thanksref{e2,addr2}}

\thankstext{e1}{e-mail: nivaldolemos@id.uff.br}
\thankstext{e2}{e-mail: reboucas.marcelo@gmail.com}

\institute{Instituto de F\'{\i}sica, Universidade Federal Fluminense,
Av. Litor\^anea, S/N, $\,$24210-340 Niter\'oi -- RJ, Brazil \label{addr1}
           \and
Centro Brasileiro de Pesquisas F\'{\i}sicas,
Rua Dr.\ Xavier Sigaud 150,
$\,$22290-180 Rio de Janeiro -- RJ, Brazil\label{addr2}
}

\date{Received: date / Accepted: date}

\maketitle

\sloppy

\begin{abstract}
Orientability is an important global topological property of spacetime
manifolds. It is often assumed that a test for spatial orientability
requires a global journey across the whole $3-$space to check for
orientation-reversing paths.
Since such a global expedition is not feasible, theoretical arguments
that combine universality of physical experiments with local arrow of time,
CP violation and CPT invariance are usually offered to support the choosing
of time- and space-orientable spacetime manifolds.
Another theoretical argument also offered to support this choice  comes
from the impossibility of having globally defined spinor fields on
non-orientable spacetime manifolds.
In this paper, we argue that it is possible to locally access spatial orientability
of Minkowski empty spacetime through  physical effects involving quantum vacuum
electromagnetic fluctuations.
We  study the motions of a charged particle and a point electric dipole subject
to these electromagnetic fluctuations in Minkowski spacetime with  orientable and
non-orientable spatial topologies.
We derive analytic expressions for a  statistical orientability indicator for both
of these point-like particles in  two inequivalent spatially flat  topologies.
For the charged particle,   we show that it is  possible to distinguish the
orientable from the non-orientable topology by contrasting the time evolution
of the orientability indicators.
This result reveals  that it is possible 
to access orientability through electromagnetic quantum vacuum
fluctuations.
However, the answer to the central question of the paper, namely how to locally
probe the orientability of Minkowski $3-$space intrinsically, comes about only
in the study of the  motions of an electric dipole.
For this point-like particle, we find that  a characteristic inversion
pattern exhibited by the curves of the orientability statistical indicator
is a signature of non-orientability.
This  result makes it clear that it is possible to locally  unveil spatial
non-orientability through the inversion pattern of curves of our orientability
indicator for a point electric dipole  under quantum vacuum electromagnetic
fluctuations.
Our findings might open the way to a conceivable experiment involving quantum
vacuum electromagnetic fluctuations to locally probe the spatial orientability
 of Minkowski empty spacetime.
\end{abstract}




\section{Introduction}\label{Intro}
The Universe is modeled as a four-dimensional differentiable manifold,
which  is a topological space with an additional differential structure
that permits to  locally define connections, metric and curvature with
which the gravitation theories are formulated.
Geometry is a local attribute that brings about curvature, whereas topology
is a global feature of a manifold related, for example, to  compactness and
orientability.
Geometry constrains but does not specify the topology.
So, topologically different manifolds can have a given geometry.%
\footnote{
Despite our present-day inability to predict the spatial topology of the Universe from
a fundamental theory, one should be able to probe it through cosmic microwave background
radiation (CMBR) or (and) stochastic primordial gravitational
waves~\cite{CosmTopReviews,Reboucas2020},
which should follow some basic detectability conditions~\cite{TopDetec}.
For recent topological constraints from CMBR data  we refer the reader to
Refs.~\cite{Vaudrevange-etal-12,Planck-2013-XXVI,Planck-2015-XVIII}. For some limits
on the circles-in-the-sky method designed for  searches of cosmic topology
through CMBR see Ref.~\cite{Gomero-Mota-Reboucas-2016}.}

Since topology antecedes geometry,  it is important to determine whether, how
and to what extent physical phenomena depend upon or are somehow affected, induced,
triggered, or even driven by a nontrivial  topology. The net role played by the spatial
topology is more clearly ascertained  in the static spatially flat
Friedmann-Lema\^{\i}tre-Robertson-Walker spacetime, whose  dynamical degrees
of freedom are frozen.
Thus, in this work we  focus on Minkowski spacetime,  whose spatial geometry
is Euclidean.

Although the topology of the spatial section, $M_3$, of Minkowski spacetime,
$\mathcal{M}_4 = \mathbb{R}\times M_3$, is usually taken to be the
simply-connected Euclidean space $\mathbb{E}^{3}$, it is a  mathematical fact
that it can also be any one of the possible $17$ topologically distinct quotient
(multiply-connected) manifolds $M_3 = \mathbb{E}^3/\Gamma$, where $\Gamma$ is a
discrete group of isometries or holonomies acting freely on the covering space
$\mathbb{E}^{3}$~\cite{Wolf67,Thurston}.
The action of $\Gamma$ tiles or tessellates the covering manifold into identical
domains or cells which are copies of what is known as fundamental polyhedron (FP)
or fundamental cell or domain (FC or FD).
On the covering manifold $\mathbb{E}^{3}$, the multiple connectedness of $M_3$
is taken into account by imposing  periodic boundary 
conditions (repeated cells) that are determined by the action of the
group of discrete isometries $\Gamma$ on the covering space $\mathbb{E}^{3}$.

In a  manifold with periodic boundary conditions only certain modes of fields
can exist. 
Thus, a nontrivial  topology may leave its mark on the expectation values
of local physical quantities. 
A  case in point is the  Casimir effect of topological origin
~\cite{dhi79,Dowker-Critchley-1976,st06,MD-2007,DHO-2002,DHO-2001}.

Quantum vacuum fluctuations of the electromagnetic field may give rise to
stochastic motions for charged test particles~\cite{sw08,bbf09,yf04,yc04,%
f05,ycw06,pf11,lrs16}.
In  Minkowski spacetime with the standard simply-connected spatial section, it is
unsettled whether such motions of test particles can happen~\cite{gour99,jaekel92}.
What is theoretically clear is that when changes of a topological nature are made
in the background $3-$space as, for example, the  
insertion of perfectly reflecting planes into the three-dimensional spatial section
of Minkowski spacetime, the resulting mean squared velocity of a charged test
particle   
is affected~\cite{sw08,yf04,yc04,f05,ycw06,pf11,lrs16}.

In a recent paper, it was shown that endowing the spatial sections of
Minkowski spacetime with a nontrivial topology influences the stochastic motions
performed by a test charged particle under vacuum fluctuations of the electromagnetic
field~\cite{Bessa-Reboucas-2020}.
Thus, either by inserting  perfectly reflecting boundaries
or equipping  $3-$space with any of the classified nontrivial topologies,
the  stochastic motions of charged particles under quantum vacuum electromagnetic
fluctuations are changed in a way that is measurable in principle.%
\footnote{In Refs.~\cite{bbf09,yf04,yc04,f05,ycw06,sw08,pf11,lrs16} one has
a nontrivial  inhomogeneous quotient orbifold space topology~\cite{Thurston}
while in Ref.~\cite{Bessa-Reboucas-2020} we have 
nontrivial flat smooth manifolds. 
We also note in this regard that the classification of three-dimensional Euclidean spaces
was first taken up in the field of crystallography%
~\cite{Feodoroff-1885,Bieberbach-1911,Bieberbach-1912}
and completed in 1934~\cite{Novacki-1934}. For a recent exposition the reader is referred
to Refs.~\cite{Adams-Shapiro01,Cipra02,Riazuelo-et-el03,Fujii-Yoshii-2011}. }

Orientability is an important global topological property of a spacetime manifold.
It is widely assumed, implicitly or explicitly, that a manifold modeling the physical
spacetime is globally orientable in all respects. Namely, that it is spacetime
orientable and, additionally, that it is separately time and space orientable.
Besides, it is also generally assumed that, being a global property, the $3-$space
orientability cannot be tested locally. Thus, to disclose the spatial 
orientability one would have to make  trips along some specific closed paths
around the whole $3-$space to check, for example, whether one returns with left-
and right-hand sides exchanged.
This reasoning is at first sight open to an immediate objection:     
since such a global journey across the whole $3-$space is 
not feasible one might think that  spatial orientability cannot be   
probed. In the face of this hurdle, one would have either to derive it from a fundamental
theory of physics or answer the orientability question through cosmological observations
or  local experiments.
Thus, it is conceivable that  spatial orientability might be subjected to
local experimental tests.%
\footnote{One can certainly take advantage of gedanken experiments to reach theoretical
conclusions, but not as a replacement to actual experimental evidence in
physics~\cite{Anandan}.}

Since  quantum vacuum fluctuations of the
electromagnetic field can be used to disclose a putative nontrivial $3-$space
topology of Minkowski spacetime through stochastic
motions of test charged particles~\cite{Bessa-Reboucas-2020}, and given that $8$
out of the possible $17$ quotient flat $3-$manifolds are non-orientable~\cite{Wolf67},
a question that naturally arises  is whether these quantum vacuum fluctuations
could be also used to reveal locally specific topological properties such as orientability
of $3-$space.%
\footnote{In mathematical terms, this amounts to identifying physical signatures of 
the flip holonomy through the local motion of point-like particles under  
electromagnetic vacuum fluctuations.}%

Our chief goal in this article is to address this question by inquiring the
electromagnetic quantum fluctuations  about the spatial orientability of
Minkowski spacetime.
To this end,  we investigate stochastic motions of a charged particle
and of an electric dipole under quantum fluctuations of the electromagnetic field in
Minkowski spacetime with two inequivalent spatial topologies, namely
the orientable slab  ($E_{16}$) and the non-orientable slab with flip ($E_{17}$).%
\footnote{In the next section we present a summary of flat three-dimensional
topologies and some of their main topological properties.
 For a more detailed account on these topologies we recommend
Refs.~\cite{Adams-Shapiro01,Cipra02,Riazuelo-et-el03,Fujii-Yoshii-2011}.}
These topologies turn out to be  
suitable to identify orientability or non-orientability
signatures by stochastic motions 
of point-like particles in Minkowski spacetime.

In Section~\ref{TopSet} we introduce the notation and present some key concepts
and results regarding topologies of three-dimensional manifolds, which will be needed
in the rest of the paper.
In Section~\ref{Syst-montion} we present the physical systems along with the
background geometry and topology,
introduce the orientability statistical indicator and derive its expressions
for both  a  charged particle and an electric dipole under quantum vacuum
electromagnetic fluctuations in Minkowski spacetime with  $E_{16}$
and $E_{17}$ flat $3-$space topologies. 
For the charged particle, we show that by comparing the time evolution
of the  orientability indicator for a particle in $E_{16}$ and $E_{17}$ one
can discriminate the orientable from the non-orientable topology.
This  result is significant in that it makes apparent the
strength of our approach to access orientability through electromagnetic
quantum vacuum fluctuations.
However, a possible answer to the central question of the paper, namely how to
locally probe the orientability of Minkowski $3-$space per se, comes about
only in the study  of the motions of an electric dipole.  
In this regard, the most important finding  is that  spatial non-orientability 
can be locally  unveiled through the inversion pattern of the curves of the
orientability statistical indicator for a point electric dipole  under quantum
vacuum electromagnetic fluctuations.
Section~\ref{Conclusion} is dedicated to our main conclusions and final
remarks.

\section{Topological prerequisites }  \label{TopSet}

Our primary aim in this section is to introduce the notation and give some basic
definitions and results concerning the topology of flat three-dimensional
manifolds that are used throughout this paper.
The spatial section $M_3$ of the Minkowski spacetime  manifold
$\mathcal{M}_4 = \mathbb{R}\times M_3$ is usually assumed to be the simply
connected Euclidean space $\mathbb{E}^{3}$.
\footnote{$\mathbb{R}^3$ is a topological space while $\mathbb{E}^3$ is a
geometrical space, i.e. $\mathbb{R}^3$ endowed with the Euclidean metric.}
But it can also be a multiply-connected quotient $3-$manifold
of the form  $M_3=\mathbb{E}^{3}/\Gamma$, where $\mathbb{E}^{3}$ is the covering
space and $\Gamma$ is a discrete and fixed-point-free group of discrete
isometries (also referred to as the holonomy group~\cite{Wolf67,Thurston})
acting freely  on the covering space $\mathbb{E}^{3}$~\cite{Wolf67}.

Possibly the best known example of three-dimensional quotient Euclidean
manifold with nontrivial topology is the $3-$torus
$T^3=\mathbb{S}^1 \times \mathbb{S}^1 \times \mathbb{S}^1=\mathbb{E}^3/\Gamma$,
whose fundamental
polyhedron (FP) is a parallelepiped with sides $a,b,c$ (say),
the opposite faces of which are identified through translations.
In any multiply-connected quotient flat $3-$manifold the fundamental polyhedron tiles
(tessellates) the whole infinite simply-connected covering space $\mathbb{E}^3$.
The group $\Gamma=\mathbb{Z} \times \mathbb{Z} \times \mathbb{Z}$ consists
of discrete translations associated with the face identification.
The periodicities in the three independent directions are given by the
circles $\mathbb{S}^1$.

In forming the quotient manifolds $M_3$ an essential  point is that  
they are obtained from the covering manifold $\mathbb{E}^3$ by identifying
points that are equivalent under the action of the discrete isometry group
$\Gamma$. Hence, each point on the quotient manifold $M_3$ represents all
the equivalent points on the covering space. The multiple connectedness
leads to periodic boundary conditions on the covering manifold $\mathbb{E}^3$
(repeated cells) that are determined by the action of the group $\Gamma$ on the
covering manifold. Clearly, different isometry groups $\Gamma$ define different
topologies for $M_3$, which in turn give rise to different periodicity
on the covering manifold (different mosaic of the covering space
$\mathbb{E}^3$).

Another important point in forming the flat quotient manifolds $M_3$ is that every
covering isometry $\gamma \in \Gamma$ can be expressed (in the covering space $\mathbb{E}^3$)
through translation, rotation, reflection (flip) and combinations thereof.
A screw motion, for example, is a combination of a rotation $R(\alpha,\mathbf{\widehat{u}})$
by an angle $\alpha$ around an axis $\mathbf{\widehat{u}}$, followed
by a translation along a vector $\mathbf{L} = L \,\mathbf{\widehat{w}}$, say.
A general glide reflection is combination of a reflection followed by a translation, as
for example  $P = (x,y,z) \,\mapsto \, \gamma P = (-x,y,z) + (0,0,c) $, 
where $c$ is a constant. If $c=0$ we have a simple reflection or flip.

In dealing with metric manifolds in mathematical physics two concepts of
homogeneity arise. Local homogeneity is a geometrical characteristic of metric
manifolds. It is formulated in terms of the action of the group  of local
isometries.
In dealing with topological spaces, we have the global homogeneity of
topological nature.
A way to characterize global homogeneity  of the quotient manifolds is
through distance functions. Indeed, for any
$\mathbf{x} \in M_3$ the distance function $\ell_\gamma (\mathbf{x})$
for a given discrete isometry $\gamma \in \Gamma$ is def\/ined by
\be
\label{dist-function}
\ell_\gamma(\mathbf{x}) = d(\mathbf{x}, \gamma \mathbf{x}) \; ,
\ee
where $d$ is the Euclidean metric defined on $M_3$. The distance function
provides the length of the closed geodesic that passes through $\mathbf{x}$
and is associated with a holonomy $\gamma$.
In  \textit{globally homogeneous} manifolds the distance function for any covering
isometry $\gamma$ is constant. In globally inhomogeneous manifolds, in contrast,
the length of the closed geodesic associated with at least one  $\gamma$
is non-translational (screw motion or flip, for example) and depends on the
point $\mathbf{x} \in M_3$, and then is not constant.

When the distance between a point $\mathbf{x}$ and its
image $\gamma \mathbf{x}$ (in the covering space) is a constant for all points
$\mathbf{x}$ then the holonomy $\gamma$ is a translation, that is,
all elements of the covering
group $\Gamma$  in globally homogeneous spaces are translations.
This means that in these manifolds the faces of the fundamental
cells are identified through independent translations.

In this paper, we shall consider the topologically nontrivial spaces $E_{16}$ and $E_{17}$.
The slab space $E_{16}$  is constructed by tessellating $\mathbb{E}^3$ by equidistant
parallel planes, so it has only one compact dimension associated with a  direction
perpendicular to those planes. Taking the $x$-direction as compact,  one has that,
with $n_x\in \mathbb{Z}$ and $a>0$, points $(x,y,z)$ and  $(x+n_xa,y,z)$ are identified
in the case of the slab space $E_{16}$.
The slab space with flip $E_{17}$ involves an additional inversion of a direction orthogonal
to the compact direction, that is, one direction in the tessellating planes is flipped
as one moves from one plane to the next. Letting the flip be in the $y$-direction,
 the identification of points  $(x,y,z)$ and $(x+n_xa,(-1)^{n_x}y,z)$ defines the
 $E_{17}$ topology.
In this way, the slab space $E_{16}$   
is globally homogeneous, whereas
the slab space with  flip, $E_{17}$, is  globally inhomogeneous since the
covering group $\Gamma$ contains a flip, which clearly is a non-translational
discrete isometry.

Orientability is another very important global (topological) property of a manifold
that measures whether one can choose consistently a clockwise orientation
for loops in the manifold.
A closed curve in a manifold $M_3$ that brings a traveler back to the starting point
mirror-reversed is called an orientation-reversing path. Manifolds that do not
have an orientation-reversing path are called \textit{orientable}, whereas manifolds
that contain an orientation-reversing path are \textit{non-orientable}~\cite{Weeks2020}.
Most surfaces that we encounter, such as cylinders,  planes and
tori are orientable, whereas the M\"obius strip and Klein bottle
are non-orientable surfaces.
For three-dimensional quotient manifolds, when the covering group $\Gamma$
contains at least one holonomy $\gamma$ that is a reflection (flip) the
corresponding quotient manifold is non-orientable.
Thus, for example, the slab space is  orientable while the slab space with
flip is non-orientable. Clearly non-orientable manifolds are necessarily
globally inhomogeneous as the covering group $\Gamma$ contains a reflection, which
obviously is a non-translational covering holonomy.

In Table~\ref{Tb-4-Orient_and_Non_orient} we collect the names and symbols
used to refer to the manifolds together with the number of compact
independent dimensions and information concerning their global homogeneity
and orientability.
In addition to the simply-connected $\mathbb{E}^{3}$ topology, these are
the three-dimensional manifolds with nontrivial topologies
that we shall be concerned with in this paper.

\begin{table}[h]
\centering
\caption{Names and symbols of the simply-connected $\mathbb{E}^{3}=E_{18}$, and two
multiply-connected flat orientable and 
non-orientable Euclidean quotient manifolds $M_3 =\mathbb{E}^3/\Gamma$ along
with the number of compact dimensions (Comp.), orientability and global (topological)
homogeneity.}
\label{Tb-4-Orient_and_Non_orient}
\begin{tabular*}{\columnwidth}{lcccc} 
\hline
Name   &   $\!\!$Symb.     &   $\!\!$Comp &  $\!\!\!$Orientable  & $\!\!\!\!\!\!$  Homogeneous \\
\hline
Slab space           &   $E_{16}$           &   $1$       &     yes                &  yes \\
Slab space with flip &   $E_{17}$           &   $1$       &     no                 &  no \\
Euclidean space      &   $E_{18}$           &   $0$       &     yes                &  yes \\
\hline
\end{tabular*}
\end{table}

Finally let us briefly mention a few results that are implicity or explicitly
used  throughout this paper (see Ref.~\cite{Geroch-Horowitz-1979} for a
detailed discussion).
Every simply-connected spacetime manifold is both time- and space-orientable.
The product of two manifolds is simply-connected if and only if each factor is.
If the spacetime is of the form $\mathcal{M}_4 = \mathbb{R}\times M_3$ then
space-orientability of the spacetime reduces to  orientability of the spatial
section $M_3$. This applies to the spacetime endowed with
the non-orientable $3-$space topology $E_{17}$ that we deal with in
this work.

Having set the stage for our investigation, in the next section we proceed to
inquire whether the topological (global) non-orientability property
of the spatial section of Minkowski spacetime manifold is amenable to be
locally  probed through the study of the motions of a charged test particle
or a point electric dipole under quantum vacuum fluctuations of the electromagnetic
field.

\section{Non-orientability from electromagnetic fluctuations}
\label{Syst-montion}

As noted before, quantum vacuum fluctuations of the electromagnetic field
give rise to stochastic motions for charged particles which are sensitive
to the topology of the background $3-$space.
In this section, we address the main underlying question of this paper, which is
whether these fluctuations offer a suitable way of discovering a putative
non-orientability of Minkowski spatial sections.    
We  take up this question through the study of stochastic motions of a charged 
particle and an electric dipole under electromagnetic quantum fluctuations in Minkowski
spacetime with two  inequivalent spatial topologies, 
namely the orientable slab space ($E_{16}$) and the non-orientable slab space with
flip ($E_{17})$.
In the following  we present the details of our investigation and main
results.

\subsection{NON-ORIENTABILITY WITH POINT CHARGED PARTICLE}
\label{Subsec-charge}

We first consider a  nonrelativistic test particle with
charge $q$ and mass $m$ locally subject to
vacuum fluctuations of the electric field ${\bf E}({\bf x}, t)$
in a topologically nontrivial spacetime manifold
equipped with the Minkowski metric $\eta_{\mu\nu}=\mbox{diag} (+1, -1, -1, -1)$.
The spatial section is usually taken to be $\mathbb{E}^{3}$,
but here we take for $M_{3}$ each of the two multiply-connected manifolds in
Table~\ref{Tb-4-Orient_and_Non_orient}.

Locally, the motion of the  charged test particle is determined by the Lorentz force.
In the nonrelativistic limit  the equation of motion for the point charge is
\begin{equation}\label{eqmotion1}
\frac{d{\bf v}}{dt} = \frac{q}{m} \,{\bf E}({\bf x}, t)\,,
\end{equation}
where  $\mathbf{v}$ is the particle's velocity and
$\mathbf{x}$ its position at  time $t$.
We assume that on the time scales of interest the particle practically does not
move, i.e. it has a negligible displacement, so we  can ignore the time
dependence of  $\mathbf{x}$.
Thus, the  particle's position $\mathbf{x}$ is taken as constant in what
follows~\cite{yf04,lrs16}.%
\footnote{The corrections arising from the inexactness of this assumption
are negligible in the low velocity regime.}
Assuming that the particle is initially  at rest,  integration of
Eq.~\eqref{eqmotion1} gives
\begin{equation}\label{eqmotion2}
{\bf v}({\bf x}, t) = \frac{q}{m}\int_0^{t}{\bf E}({\bf x}, t^{\prime})\,dt^{\prime} \,,
\end{equation}
and the mean squared velocity, velocity dispersion or simply dispersion in each of
the three independent directions $i = x, y, z$ is given by%
\footnote{By definition,
$\,\bigl \langle \,\Delta v^2_i(\mathbf{x}, t) \,\bigr \rangle =
\bigl \langle \, v^2_i(\mathbf{x}, t)  \,\bigr \rangle
-\bigl \langle \,v_i(\mathbf{x}, t)   \bigr \rangle^2$.}
\begin{equation}\label{eqdispersion1}
\Bigl \langle\Delta v^2_i \Bigr\rangle = \frac{q^2}{m^2} \int_0^t\int_0^t
\Bigl\langle E_i({\bf x}, t') E_i({\bf x}, t'')\Bigr \rangle\, dt' dt''\,.
\end{equation}
Following Yu and Ford~\cite{yf04}, we assume that the electric field is
a sum of classical  $\mathbf{E}_c$ and quantum  $\mathbf{E}_q$ parts.
Because ${\bf E}_c$ is not subject to quantum fluctuations and
$\langle {\bf E}_q\rangle =0$,
the two-point function $\langle E_i({\bf x}, t)E_i({\bf x}', t')\rangle$ in
equation~\eqref{eqdispersion1} involves only
the quantum part of the electric field~\cite{yf04}.

It can be shown~\cite{bd82} that locally
\begin{eqnarray} \label{eqdif-0}
&\Bigl \langle E_i({\bf x}, t)E_i({\bf x}', t') \Bigr \rangle  = &
\frac{\partial }{\partial x_i} \frac{\partial}
{\partial {x'}_i}D({\bf x}, t; {\bf x}', t')  \nonumber\\
 & &  - \,\, \frac{\partial }{\partial t}
 \frac{\partial}
{\partial t'}D ({\bf x }, t; {\bf x'}, t')
\end{eqnarray}
where, in Minkowski spacetime 
with $M_3= \mathbb{E}^3= E_{18}$,
the Hadamard function $D({\bf x}, t;{\bf x}', t')$ is given by
\begin{equation}\label{eqren}
D_0({\bf x}, t; {\bf x}', t') = \frac{1}{4\pi^2(\Delta t^2 - |\Delta \mathbf{x}|^2)} \,.
\end{equation}
The subscript $0$ indicates standard Minkowski spacetime,
$\Delta t = t - t'$ and  $|\Delta \mathbf{x}| \equiv r $  
is the spatial separation for topologically trivial Minkowski spacetime:
\begin{equation}\label{separation-trivial}
r^2 = (x-x')^2 + (y-y')^2 + (z - z')^2  \,.
\end{equation}


In Minkowski spacetime with a topologically nontrivial spatial section,
the {spatial separation} $r^2$ takes a different form that captures the
periodic boundary conditions imposed on the covering space $\mathbb{E}^{3}$
by the covering group  $\Gamma$, which characterize the spatial topology.
In consonance with Ref.~\cite{st06}, in Table~\ref{Tb-Spatial-separation}
we collect the {spatial separations} for the topologically inequivalent
Euclidean spaces we shall address in this paper.%
\footnote{The reader is referred to Refs.~\cite{Cipra02,Riazuelo-et-el03,%
Fujii-Yoshii-2011} for pictures of the fundamental cells and further properties of
all possible  three-dimensional Euclidean topologies.}

\begin{table}[htb]
\centering
\caption{Spatial separation in Hadamard function for  the simply-connected
Euclidean manifold $E_{18}$, and for the  multiply-connected flat
orientable ($E_{16}$) and its non-orientable counterpart ($E_{17}$) quotient
Euclidean manifolds. The topological compact length is denoted by $a$.
The numbers $n_x$ are integers  and run from $-\infty$ to $\infty$.
For each multiply-connected topology, when  $n_x=0$ we recover the
spatial separation for the simply-connected Euclidean $3-$space.}
\label{Tb-Spatial-separation}
\begin{tabular*}{\columnwidth}{ll} 
\hline
\hspace*{-.3cm} Spatial topology       & \hspace{-.4cm} Spatial separation $ r^2$  for Hadamard function  \\
\hline
\hspace*{-.3cm} $E_{16}$ -  Slab space           &  $(x - x'- n_x a)^2 + (y - y')^2 + (z - z')^2 $ \\
\hline
\hspace*{-.3cm} $E_{17}$ -  Slab space with flip &  \hspace{-.32cm} $\left(x-x'-n_x a \right)^{2}+\left(y -(-1)^{n_x} y'\right)^{2}
+ \left(z - z' \right)^{2}$ \\
 \hline
\hspace*{-.3cm} $E_{18}$ -  Euclidean space      & \hspace{.3cm} $(x - x')^2 + (y - y')^2 + (z - z')^2 $\\
\hline
\end{tabular*}
\end{table}


\vspace{4mm}
\noindent 
\textbf{\small ORIENTABILITY INDICATOR -- $\,$SLAB SPACE WITH FLIP $\mathbf{E_{17}}$}
\vspace{1mm}

For the sake of brevity, we present  detailed calculations only
for a charged particle in Minkowski spacetime with $E_{17}$ spatial topology.
The corresponding results for $E_{16}$  can then
be easily obtained from those for $E_{17}$, as we show below.

To obtain the correlation function for the electric field
that is required to compute the velocity dispersion~\eqref{eqdispersion1} for
slab space with flip $E_{17}$,  we replace in Eq.~(\ref{eqdif-0}) the Hadamard
function $D({\bf x}, t; {\bf x}', t')$ by its renormalized version
given by~\cite{Bessa-Reboucas-2020}
\begin{eqnarray}\label{Hadamard-ren}
D_{ren}({\bf x}, t; {\bf x}', t') & = &
D({\bf x}, t; {\bf x}', t') - D_0({\bf x}, t; {\bf x}', t')\nonumber \\
& = &
\sum\limits_{{n_x=-\infty}}^{{\infty\;\;\prime}}\frac{1}{4\pi^2(\Delta t^2 - r^2)}\,,
\end{eqnarray}
where  here and in what follows $\sum_{}^{\;'}$ indicates that
the Minkowski contribution term $n_x = 0$ is excluded from the summation,
$\Delta t = t-t^{\prime}$, and, from Table~\ref{Tb-Spatial-separation},
the spatial separation for $E_{17}$ is
\begin{equation}\label{separation-E17}
r^2 = \left( x - x' - n_x a \right)^{2} +\left(  y -(-1)^{n_x} y'\right)^{2}
+ \left(z - z' \right)^{2}.
\end{equation}
The term  with $n_x=0$ in  the sum~\eqref{Hadamard-ren} is the Hadamard function
$D_{0}({\bf x}, t; {\bf x}', t')$ for  Minkowski spacetime with simply-connected
spatial section $E_{18}$.
This term  has been subtracted out from the sum
because it gives rise to an infinite contribution
to the velocity dispersion.  

Thus, from equation \eqref{eqdif-0} the renormalized correlation functions
\begin{eqnarray}\label{correlation-i-E17}
\bigl \langle E_i({\bf x}, t)E_i({\bf x}', t')\bigr \rangle_{ren} & = &
\frac{\partial }{\partial x_i} \frac{\partial}
{\partial {x'}_i}D_{ren}({\bf x}, t; {\bf x}', t') \nonumber \\
& & - \frac{\partial }{\partial t} \frac{\partial}
{\partial t'}D_{ren} ({\bf x }, t; {\bf x'}, t')
\end{eqnarray}
are then given by
\begin{eqnarray}\label{correlation-x-E17}
\bigl \langle E_x({\bf x}, t)E_x({\bf x}', t')\bigr \rangle_{ren}^{E_{17}} &  = &
\sum\limits_{{n_x=-\infty}}^{{\infty\;\;\prime}}
\frac{\Delta t^2 + r^2 -2r_x^2}{\pi^2 [\Delta t^2 - r^2]^3},\\
\label{correlation-y-E17}
\bigl \langle E_y({\bf x}, t)E_y({\bf x}', t')\bigr \rangle_{ren}^{E_{17}} & = &
\sum\limits_{{n_x=-\infty}}^{{\infty\;\;\prime}}  \bigg\{
\frac{\left( 3-(-1)^{n_x}\right) \Delta t^2 } {2\pi^2 [\Delta t^2 - r^2]^3} \nonumber \\
& &  \hspace{-1cm} +
\frac{ \left(1+(-1)^{n_x}\right) r^2-4(-1)^{n_x} r_y^2} {2\pi^2 [\Delta t^2 - r^2]^3}\bigg\}, \\
\label{correlation-z-E17}
\bigl \langle E_z({\bf x}, t)E_z({\bf x}', t')\bigr \rangle_{ren}^{E_{17}} &  = &
\sum\limits_{{n_x=-\infty}}^{{\infty\;\;\prime}}
\frac{\Delta t^2 + r^2 -2r_z^2}{\pi^2 [\Delta t^2 - r^2]^3},
\end{eqnarray}
where $\Delta t= t -t^{\prime}$ and
\begin{eqnarray}\label{r-components-E17}
r_x & = & x-x^{\prime} - n_xa, \qquad r_y = y-(-1)^{n_x} y^{\prime}, \nonumber\\
 r_z & = & z-z^{\prime}, \hspace{1.6cm}  r= \sqrt{r_x^2 + r_y^2+ r_z^2}.
\end{eqnarray}

The orientability indicator $\mbox{\large $I$}_{v^2_i}^{E_{17}}$ that
we will consider is defined by replacing
the electric field correlation functions in Eq.~(\ref{eqdispersion1}) by
their renormalized counterparts~(\ref{Hadamard-ren}) in which $r$ is given
by~(\ref{separation-E17}):
\begin{equation}   \label{indicator-E17}
\mbox{\large $I$}_{v^2_i}^{E_{17}} ({\bf x},t) = \frac{q^2}{m^2} \int_0^t\int_0^t
\Bigl\langle E_i({\bf x}, t') E_i({\bf x}, t'')\Bigr \rangle_{ren}^{E_{17}}\, dt' dt''\,.
\end{equation}
{}From~(\ref{Hadamard-ren}) it is clear  that the orientability indicator
$\mbox{\large $I$}_{v^2_i}^{E_{17}}$ is the difference between the velocity
dispersion in $E_{17}$ and the one in Minkowski with trivial (simply connected)
topology.

Before proceeding to the calculation of the components of the orientability
indicator~\eqref{indicator-E17} some words of clarification are in order
for the sake of generality and for later use. 
From equations~\eqref{eqdispersion1} and~\eqref{Hadamard-ren}
a general definition of the orientability indicator can be written in the form
\begin{equation}  \label{new-ind}     
 \mbox{\large $I$}_{v^2_i}^{MC}
= \Bigl\langle\Delta v_i^2 \Bigr \rangle^{MC}
- \;\,\Bigl \langle\Delta v_i^2 \Bigr \rangle^{SC} ,
\end{equation}
where  $\bigl\langle\Delta v_i^2 \bigr \rangle$  is the
mean square velocity dispersion, and the superscripts $MC$ and $SC$
stand for multiply- and simply-connected manifolds, respectively.
The right-hand side of~(\ref{new-ind}) is defined by first taking the difference
of the two terms with ${\bf x}^{\prime} \neq  {\bf x}$ and then setting
${\bf x}^{\prime} =  {\bf x}$. Since $\mbox{\large $I$}_{v^2_i}^{MC}$ is not 
the velocity dispersion $\bigl\langle\Delta v_i^2 \bigr\rangle^{MC}$ but the
difference \eqref{new-ind}, the possibility that it takes negative values
should not be cause of concern,%
\footnote{It is neither experimentally nor theoretically  settled  whether
the second term on the right-hand side of equation~(\ref{new-ind}) vanishes or not.
Here we take the more general view that it is nonzero.
It is only when the rather particular assumption that it vanishes is made  that
one encounters counterintuitive negative values for mean square velocities often
found in the literature~\cite{yf04,Bessa-Reboucas-2020,yc04,f05,ycw06,sw08,pf11,%
lrs16,Chen}. This may be looked upon as an indication that the simply-connected
term in equation~(\ref{new-ind}) should not vanish.}
a point that does not seem to have been  appreciated in some previous works in which
this indicator was implicity used~\cite{yf04,Bessa-Reboucas-2020,yc04,%
f05,ycw06,sw08,pf11,lrs16,Chen} together with the particular assumption that
the second term vanishes.
A similar statistical indicator that measures the departure of a statistical
quantity from its values for the simply-connected space
comes about in cosmic crystallography, which is an  approach to detect cosmic
topology from the distribution of  discrete cosmic sources~\cite{LeLaLu}.
Indeed, a topological signature of any multiply connected
$3-$manifold of constant curvature is given by a constant times the
dif\/ference $\Phi_{exp}^{MC}(s_i) - \Phi^{SC}_{exp}(s_i)$ of
the expected pair separation histogram (EPSH) corresponding
to the multiply connected manifold minus the EPSH for the
underlying simply connected covering manifold~\cite{GRT01,GTRB98},
whose expression can be derived in an analytical form~\cite{GRT01,Reboucas-2000}.%

We shall discuss  experimental features of the orientability
indicator~\eqref{new-ind} in the two closing paragraphs of Subsection \ref{Dipole-motion}
in this section.
But now let us return to the  calculation of the 
components of the orientability indicator $\mbox{\large $I$}_{v^2_i}^{E_{17}}$.
Following Ref.~\cite{Bessa-Reboucas-2020},  they can  be computed with
the help of the integrals   
\begin{eqnarray}\label{integral1}
{\cal I} & = &\int_0^t  \int_0^t dt^{\prime}dt^{\prime\prime} \frac{1}{(\Delta t^2 -r^2)^3} \nonumber \\
& = & \frac{t}{16r^5 (t^2-r^2)}\bigg\{ 4rt - 3(r^2-t^2) \ln \frac{(r-t)^2}{(r+t)^2} \bigg\}
\end{eqnarray}
and
\begin{eqnarray}\label{integral2}
{\cal J} & = &\int_0^t  \int_0^t dt^{\prime} dt^{\prime\prime}
\frac{\Delta t^2}{(\Delta t^2 -r^2)^3}  \nonumber \\
& = &\frac{t}{16r^3(t^2-r^2)}\bigg\{ 4rt + (r^2-t^2)
\ln  \frac{(r-t)^2}{(r+t)^2} \bigg\},
\end{eqnarray}
in which $\Delta t = t^{\prime} -t^{\prime\prime}$.

By using these integrals and  equations~(\ref{correlation-x-E17})
to~(\ref{r-components-E17}) with  ${\bf x}^{\prime} = {\bf x}$ in
Eq.~(\ref{indicator-E17}) we find
\begin{eqnarray}\label{dispersion-x-E17}
\mbox{\large $I$}_{v^2_x}^{E_{17}}({\bf x},t) &\!\!  = \!\! \!\! &
\sum\limits_{{n_x=-\infty}}^{{\infty\;\;\prime}}\!\!  \frac{q^2 t}{16\pi^2 m^2 r^5 (t^2 - r^2)}
\bigg\{ 4rt({\bar r}_x^{\,\, 2} + r^2) \nonumber \\
& & + (t^2-r^2)(3{\bar r}_x^{\,\, 2} -r^2) \ln \frac{(r-t)^2}{(r+t)^2} \bigg\}, \\
\label{dispersion-y-E17}
\mbox{\large $I$}_{v^2_y}^{E_{17}}({\bf x},t) &\!\!  = \!\! \!\! &
\sum\limits_{{n_x=-\infty}}^{{\infty\;\;\prime}} \frac{q^2 t}{32\pi^2 m^2 r^5 (t^2 - r^2)}
\bigg\{ 4rt{\bar r}_y^{\,\, 2} \nonumber \\
& & \hspace{-.35cm} + 4rt(3-(-1)^{n_x})r^2 \nonumber \\
& & \hspace{-.35cm} + (t^2-r^2)\bigl[3{\bar r}_y^{\,\, 2} -(3-(-1)^{n_x})r^2\bigr]
\ln \frac{(r-t)^2}{(r+t)^2} \bigg\}, \\
\label{dispersion-z-E17}
\mbox{\large $I$}_{v^2_z}^{E_{17}}({\bf x},t) &\!\!  = \!\! \!\! &
\sum\limits_{{n_x=-\infty}}^{{\infty\;\;\prime}}\!\!  \frac{q^2 t}{16\pi^2 m^2 r^5 (t^2 - r^2)}
\bigg\{ 4rt({\bar r}_z^{\,\, 2} + r^2) \nonumber \\
& & +  (t^2-r^2)(3{\bar r}_z^2 -r^2) \ln \frac{(r-t)^2}{(r+t)^2} \bigg\},
\end{eqnarray}
where
\begin{eqnarray}\label{r-ri-coincidence-r}
r & = &  r_{n_x} = \sqrt{n_x^2a^2+2(1-(-1)^{n_x})y^2}, \\
\label{r-ri-coincidence-r-bar-x}
{\bar r}_x^2 & = & r^2- 2r_x^2 = -n_x^2 a^2 + 2(1-(-1)^{n_x})y^2, \\
\label{r-ri-coincidence-r-bar-y}
{\bar r}_y^2 & = & (1+ (-1)^{n_x}) r^2 - 8 (-1)^{n_x} (1-(-1)^{n_x})y^2 \nonumber \\
&  = &  (1+ (-1)^{n_x})n_x^2 a^2 + 8 (1-(-1)^{n_x})y^2,\\
\label{r-ri-coincidence-r-bar-z}
{\bar r}_z^2 & = & r^2- 2r_z^2 = r^2\,,
\end{eqnarray}
as follows from equations~(\ref{r-components-E17}) in the coincidence limit.

Before proceeding to $E_{16}$ topology, as a consistency check of the above
calculations we discuss the topological Minkowskian limit for the orientability
indicator $\mbox{\large $I$}_{v^2_i}^{E_{17}}$.
We begin by recalling
that compact lengths associated with Euclidean quotient manifolds are not fixed.
Different values of $a$  correspond to different $3-$manifolds with the same topology.
Intuitively, by letting the compact length $a \to \infty$ the role of the compactness
(nontrivial topology) should disappear and a topological Minkowskian simply-connected
limit for the orientability indicator should be attained, which is zero by
definition~(\ref{new-ind}).
In order to verify if this is the case, we first note that from Eq.~(\ref{r-ri-coincidence-r})
it follows that letting $a \to \infty$ amounts to letting $r \to \infty$.
For very large $r$ each term of the sum~(\ref{dispersion-x-E17})
consists of a fraction whose numerator is dominated by a power of $r$ not bigger than the
fourth (the logarithmic term tends to zero  as $r \to \infty$) whereas the denominator
becomes proportional to $r^7$. Therefore each term of the sum vanishes in the limit
$a \to \infty$ and $\mbox{\large $I$}_{v^2_x}^{E_{17}}$  is zero. 
The same argument shows that the other components  of the orientability indicator 
also vanish in the limit $a \to \infty$, in agreement with (\ref{new-ind}).

In the  opposite limit of small $a$ the effects of the topology become arbitrarily
large: the indicators~(\ref{dispersion-x-E17}) to (\ref{dispersion-z-E17})
grow proportionally to $1/a^2$ as $a \to 0$, and the same is true regarding
the indicators for $E_{16}$ given by equations~(\ref{dispersion-orientable-x-E16})
and~(\ref{dispersion-orientable-yz-E16}) below. This behavior is intuitively expected
since for  $a \ll t$ in the time interval  of interest an electromagnetic signal
would cross an enormous number of fundamental cells, and the effect of the
nontrivial topology would be huge.

\vspace{2mm}
\noindent
\textbf{\small ORIENTABILITY INDICATOR \  -- \  SLAB SPACE $\mathbf{E_{16}}$}
\vspace{2mm}

The factors of $(-1)^{n_x}$ that appear in equations~(\ref{correlation-y-E17})
and~(\ref{dispersion-y-E17}) arise from derivatives with respect to~$y^{\prime}$
in Eq.~(\ref{correlation-i-E17}) contributed by the separation $r$ given by
Eq.~(\ref{separation-E17}). Hence, the results for $E_{16}$ are immediately
obtained from those for $E_{17}$ by simply replacing $(-1)^{n_x}$ by $1$ everywhere
in Eqs.~(\ref{dispersion-x-E17}) to (\ref{r-ri-coincidence-r-bar-z}).
This leads to
\begin{eqnarray}\label{dispersion-orientable-x-E16}
\mbox{\large $I$}_{v^2_x}^{E_{16}}({\bf x},t) &=&
-\frac{q^2 t}{4\pi^2 m^2}\sum\limits_{{n_x=-\infty}}^{{\infty\;\;\prime}}\!\!
\frac{1}{n_x^3a^3} \ln \frac{(n_xa-t)^2}{(n_xa+t)^2},\\
\label{dispersion-orientable-yz-E16}
\mbox{\large $I$}_{v^2_y}^{E_{16}}({\bf x},t) &=&
\mbox{\large $I$}_{v^2_z}^{E_{16}}({\bf x},t)
= \frac{q^2 t}{8\pi^2 m^2} \sum\limits_{{n_x=-\infty}}^{{\infty\;\;\prime}}
\!\! \bigg\{ \frac{4t}{n_x^2a^2(t^2- n_x^2 a^2)} \nonumber \\
& & \hspace{2.8cm} + \frac{1}{n_x^3a^3} \ln \frac{(n_xa-t)^2}{(n_xa+t)^2} \bigg\}\,,
\end{eqnarray}
in agreement with the results obtained in Refs.~\cite{Chen,Bessa-Reboucas-2020},
where the indicator~\eqref{new-ind} was used.

\vspace{5mm}
\noindent
\textbf{\small NON-ORIENTABILITY WITH A CHARGED PARTICLE -- CONCLUSIONS}
\vspace{2mm}

The  components  of the orientability indicator are singular at $t=r=r_{n_x}$,
where $r_{n_x}=\vert n_x\vert a\,$
for $E_{16}$ while   $r_{n_x}$ is   given by
Eq.~(\ref{r-ri-coincidence-r}) for $E_{17}$. In both cases, the  singularities are of
topological origin and correspond to  each of the infinitely many
distances $r_{n_x}$ that  arise  from the periodic boundary conditions imposed by the
topologies on the covering space $\mathbb{E}^3$, which are given by the
identifications $\, (x,y,z) \leftrightarrow (x + n_x a, y, z) \,$
and $\, (x,y,z) \leftrightarrow (x+ n_x a,(-1)^{n_x}y,z) \,$ for $E_{16}$
and $E_{17}$, respectively.
Incidentally, we note that for a quotient orbifold
arising from the action of $\mathbb{Z}_2$ on $\mathbb{E}^3$ by reflection%
~\cite{Thurston}, which is a topological space resulting from the insertion
of a perfectly reflecting plane often used in the literature~\cite{yf04,Chen%
,sw08,ycw06,lrs16}, the  indicator~\eqref{new-ind} exhibits only one
singularity as should be expected from its origin, i.e. from the orbifold
topology properties~\cite{Thurston,Chen,Bessa-Reboucas-2020}.
This singularity has been ascribed to  local properties of the physical
system, and functions of time to switch on and off the interaction between
the particle and the electromagnetic field have been suggested to
regularize it~\cite{sw08,lrs16,lr19}.
However, this procedure does not
seem appropriate to cope with singularities of  topological nature as 
the one provided by the reflection orbifold.   
On the other hand, it is also possible  that in being of topological
origin these singularities are not only unavoidable but also unremovable in the
sense that one can remove them only by `switching off' the topology. 

\begin{figure*}[tb]           
\begin{center} 
\includegraphics[width=7.1cm,height=5.9cm]{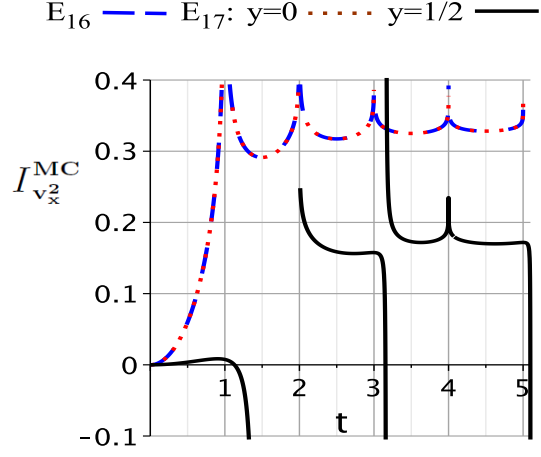} 
 \hspace{8mm}  %
\begin{minipage}[t]{\textwidth} \vspace{-1mm}
\renewcommand{\baselinestretch}{0.96}
\caption{Time evolution of the $x$-component of the  orientability indicator
$\mbox{\large $I$}_{v^2_x}^{MC}({\bf x},t)$ in units of $q^2/m^2$
for a test particle with mass $m$ and charge $q$ in Minkowski spacetime
with spatial section endowed with the non-orientable and global
inhomogeneous $E_{17}$ and orientable $E_{16}$ topologies, both with compact
length $a=1$. We show one curve for the globally homogeneous $E_{16}$
(dashed line) and two curves for $E_{17}$:  dotted and  solid lines,  
for the  particle  at the positions $P_0 = (x,0,z)$ and
$P = (x,1/2,z)$, respectively. The figure illustrates the topological
inhomogeneity of $E_{17}$, and shows that when the degree of inhomogeneity is frozen
the indicator curves for $E_{17}$ [for the particle at  $P_0 = (x,0,z)$]   
and $E_{16}$ [for the particle at generic $P=(x,y,z)\,$] coincide.
\label{E16_vs_E17-fig1}  }
\end{minipage}
\end{center}
\end{figure*}

An important question that arises at this point is what then we ultimately
learn from the above calculations regarding the local test of spatial
orientability by studying the motions of a charged particle under
electromagnetic quantum fluctuations.
In other words, what these  fluctuations are teaching us about
orientability,
equations~\eqref{dispersion-x-E17}~--~\eqref{dispersion-z-E17}
for the  non-orientable $E_{17}$ topology and equations~\eqref{dispersion-orientable-x-E16}~--~\eqref{dispersion-orientable-yz-E16}
for the orientable  $E_{16}$ space topology.
Let us now discuss this capital issue.
Clearly, conclusions can only be reached by comparisons between the
stochastic motions of the charged test particles lying in space manifolds
with each of the two topologies. 
In this regard, a first difficulty one encounters is how to make a proper
comparison because $E_{16}$ is globally homogenous whereas $E_{17}$ is not
(cf.~Table~\ref{Tb-4-Orient_and_Non_orient}).
This means that the orientability indicator  does not depend on the
particle's position for $E_{16}$, but it does when the particle lies
in a space with the globally inhomogeneous topology $E_{17}$.
The functional dependence of the dispersion on the particle's position coordinates
in these manifolds makes apparent the first difficulty. Indeed,  the components of
the orientability indicator~\eqref{dispersion-x-E17}~--~\eqref{dispersion-z-E17}
for $E_{17}$ depend on the $y$-coordinate,  while the components  ~\eqref{dispersion-orientable-x-E16}~--~\eqref{dispersion-orientable-yz-E16} for
$E_{16}$ do not.
Thus, one has to suitably choose the point $P = (x,y,z)$ in  $E_{17}$
for the  particle's position in order to make a proper comparison
between the orientability indicator curves for the topologically homogeneous $E_{16}$
and the  topologically inhomogeneous $E_{17}$ manifolds.
{}From the identification of $(x,y,z)$ and $(x+n_xa,(-1)^{n_x}y,z)$
that defines the $E_{17}$ topology, clearly a suitable way to freeze out
the global inhomogeneity degree of freedom, and thus isolate the non-orientability
effect, is by choosing as the  particle's position the point $P_0 = (x,0,z)$.
Since our chief concern is orientability,  in all figures in this paper
but one (Fig.~\ref{E16_vs_E17-fig1}) we choose this point as the particle's position
when dealing with $E_{17}$ topology.

\begin{figure*}[bt]           
\begin{center} 
\includegraphics[width=7.1cm,height=5.9cm]{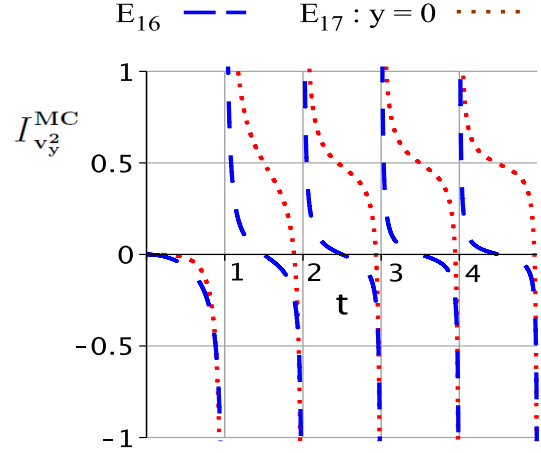} 
 \hspace{8mm}  %
\begin{minipage}[t]{\textwidth} \vspace{-1mm}
\renewcommand{\baselinestretch}{0.96}
\caption{Time evolution of the $y$-component of the orientability indicator
$\mbox{\large $I$}_{v^2_y}^{MC}$ in units of $q^2/m^2$
for a charged test particle  in Minkowski spacetime
with spatial section endowed with the orientable $E_{16}$ and non-orientable $E_{17}$
topologies, both with compact length $a=1$.
We show curves for $E_{16}$ (dashed line) and for $E_{17}$
with the particle at $P_0= (x,0,z)$ (dotted line).
The indicator curves exhibit similar periodic  patterns   
for time intervals of the order of one, but it is  possible to distinguish the
two topologies by contrasting
the $y$-component of their orientability indicator.
\label{E16_vs_E17-fig2}  }  \vspace{-5mm}  
\end{minipage}
\end{center}
\end{figure*}

Having circumvented  this particle position difficulty related to the topological
inhomogeneity of $E_{17}$,
in  Fig.~\ref{E16_vs_E17-fig1} we illustrate
how global inhomogeneity affects the time evolution of the  orientability
indicator~(\ref{dispersion-x-E17}).
For two particle's positions, one  with $y=0$ and another with $y=1/2$ for compact
length $a=1$, in $E_{17}$ there arise
different curves. This shows that the indicator~\eqref{new-ind}  is able to capture
the non-homogeneity of topological origin as has been pointed out in~\cite{Bessa-Reboucas-2020}.
To allow reproduction, we mention that Figures~\ref{E16_vs_E17-fig1}
and~\ref{E16_vs_E17-fig2}  arise from Eqs.~(\ref{dispersion-x-E17})~%
--~(\ref{r-ri-coincidence-r-bar-z})
as well as~(\ref{dispersion-orientable-x-E16}) and (\ref{dispersion-orientable-yz-E16})
with compact length $a=1$ and $n_x \neq 0$ ranging from $-50$ to $50$.

Figure~\ref{E16_vs_E17-fig1} also shows that for the particle's position
$P_0$  in  $E_{17}$ the  orientability indicator curves coincide with 
those for a generic point in globally homogenous $E_{16}$.
This shows that for 
$P_0 = (x,0,z)$, where the global inhomogeneity degree of freedom
is frozen, the $x$-component of the orientability indicator cannot be
used to distinguish between the two orientable and non-orientable $3-$spaces.
With different pattern of curves,  the same  holds true for
the $z$-components of the orientability indicator  for $E_{17}$ and $E_{16}$,
but we do not show this figure  for the sake of brevity.
This result  on these two components  of the orientability indicator can be read
from our equations.
Indeed, for the particle's  position at $P_0 = (x,0,z)$ for $E_{17}$, taking
account of \eqref{r-ri-coincidence-r}~--~\eqref{r-ri-coincidence-r-bar-z}
it is straightforward to show that
Eqs.~\eqref{dispersion-x-E17} and \eqref{dispersion-z-E17}
reduce, respectively,  to \eqref{dispersion-orientable-x-E16}
and~\eqref{dispersion-orientable-yz-E16}.
However, it should be noticed that even for $P_0 = (x,0,z)$ the
$y$-component of the orientability indicator
 $\mbox{\large $I$}_{v^2_y}^{E_{17}} $
does not reduce to
$\mbox{\large $I$}_{v^2_y}^{E_{16}}  $.
This means that in order  to extract information regarding orientability from the
motion of a charged particle under electromagnetic fluctuations
the proper comparison should be between the $y$-components of the dispersion, as
we do in Fig.~\ref{E16_vs_E17-fig2}.
This was to  be expected from the outset since the  reflection holonomy for $E_{17}$
is in the $y$-direction (cf.\ Table~\ref{Tb-Spatial-separation}).

Figure~\ref{E16_vs_E17-fig2} displays the $y$-components of the orientability
indicator for a charged test particle in Minkowski space whose spatial section
has ${E_{16}}$ (orientable) and ${E_{17}}$ [non-orientable, $P_0=(x,0,z)$]
topologies.
The chief conclusion is that the component along the direction of the
flip (cf. Table~\ref{Tb-Spatial-separation}) might be used to find out whether
the particle lies in Minkowski spacetime with  orientable or  non-orientable
$3-$space. Figure~\ref{E16_vs_E17-fig2} also shows different orientability
indicator curves for $E_{16}$ and $E_{17}$ which, in both cases, repeat
themselves periodically. For the two topologies the overall patterns of the
orientability indicator curves are similar.

Although the above result seems valuable
to the extent that it makes clear the strength of our approach
to access orientability through  electromagnetic
vacuum fluctuations, it demands a comparison between the orientability indicator curves
for the two given spatial manifolds for a verdict about orientability.  Thus,
it does not provide a conclusive answer to the central question of this paper, namely
how to locally probe the orientability of the spatial section of Minkowski spacetime
in itself from electromagnetic vacuum fluctuations. 
Given the directional properties of a point electric dipole, a pertinent question 
that emerges here is whether it would be more effective to use it for 
testing the individual non-orientability of a generic $3-$space.
If so  we could  envisage a local experiment to probe the orientability of a $3-$space
per se.
In the next subsection we shall investigate  this remarkable possibility.

\subsection{NON-ORIENTABILITY WITH POINT ELECTRIC DIPOLE} \label{Dipole-motion}

A noteworthy  outcome of the previous section is that
the time evolution of the velocity dispersion for a charged particle
can be  used to locally differentiate  an orientable ($E_{16}$) from a
non-orientable ($E_{17}$) spatial section of Minkowski spacetime.
However,  it cannot be used to decide whether a given $3-$space manifold
is or not orientable. In this way, it cannot be taken as a definite answer
to our central question about the spatial orientability of Minkowski spacetime.
So, a question that naturally arises here is whether the velocity dispersion
of a different type of point-like particle could provide a suitable local
signature of non-orientability.
As a point electric dipole has directional properties, one might expect
that its orientability indicator would potentially be more effective in providing
information regarding non-orientability associated with a particular direction.
 To examine this issue  we now turn our attention 
to topologically induced motions of an electric dipole under quantum vacuum
electromagnetic fluctuations.

Newton's second law for a point electric dipole of mass $m$ in an external electric
field reads
\begin{equation}
\label{eqmotion-dipole}
m\frac{d{\bf v}}{dt} = {\bf p}\cdot {\mbox{\boldmath $\nabla$}}{\bf E}({\bf x},t)\,,
\end{equation}
where $\bf p$ is the electric dipole moment.
With the same hypotheses as for the point charge and assuming the dipole is initially
at rest,  integration of Eq.~\eqref{eqmotion-dipole} yields
\begin{equation}\label{eqmotion-dipole-integrated}
{\bf v}({\bf x}, t) = \frac{1}{m}p_j\int_0^{t}\partial_j{\bf E}({\bf x}, t^{\prime})\,dt^{\prime}
\end{equation}
with $\partial_j = \partial/\partial x_j$ and summation over repeated indices implied.

The mean squared speed in each of the three independent directions
$i = x, y, z$ is given by
\begin{equation}\label{eqdispersion-dipole}
\Bigl \langle\Delta v^2_i\Bigr \rangle = \frac{p_jp_k}{m^2} \int_0^t\int_0^t
\Bigl \langle \bigl(\partial_j E_i({\bf x}, t')\bigr)
\bigl(\partial_k E_i({\bf x}, t'')\bigr) \Bigr \rangle\, dt' dt''\,,
\end{equation}
which can be conveniently rewritten as
\begin{equation}\label{eqdispersion-dipole-rewritten}
\Bigl \langle\Delta v^2_i\Bigr \rangle = \lim_{{\bf x}^{\prime} \to {\bf x}}\frac{p_jp_k}{m^2}
\int_0^t\int_0^t \partial_j\partial_k^{\,\prime} \Bigl \langle  E_i({\bf x}, t')
 E_i({\bf x}', t'')\Bigr \rangle\, dt' dt''
\end{equation}
where $\partial_k^{\,\prime}= \partial/\partial x_k^{\prime}$.

Now we proceed to the computation of  the velocity dispersion for a point dipole
in spaces $E_{17}$ and $E_{16}$. The space $E_{17}$ has two topologically
conspicuous directions:    
the compact $x$-direction  and the flip $y$-direction associated
with the non-orientability of $E_{17}$.
To probe the non-orientability of $E{_{17}}$ by means
of stochastic motions, it seems most promising to choose  a dipole oriented  %
in the $y$-direction, since the orientation of the dipole would also be flipped
upon every displacement by the topological length  $a$ along the compact direction.
Indeed, it is for a dipole oriented in the flip direction that the effect  of
the non-orientability is most noticeable, as we show in the following.

For a dipole oriented along the $y-$axis the dipole moment is ${\bf p}=(0,p,0)$ and
we again use the indicator in which the electric field correlation functions are
replaced by their renormalized counterparts, as in equation~\eqref{indicator-E17}
for the charged point particle.
We have
\begin{eqnarray}\label{eqdispersion-dipole-y}
& & ^{^{(y)}}{\!} \mbox{\large $I$}_{v^2_i}^{E_{17}}({\bf x},t) =  \nonumber \\
& & \hspace{.5cm} \lim_{{\bf x}^{\prime} \to {\bf x}}\frac{p^2}{m^2}
\int_0^t\int_0^t \partial_y\partial_{y^{\prime}} \Bigl \langle  E_i({\bf x}, t')
 E_i({\bf x}', t'')\Bigr \rangle_{ren}^{E_{17}}\, dt' dt'' \,,
\end{eqnarray}
where the left superscript within parentheses  indicates the dipole's orientation.
With the help of Eq. (\ref{correlation-x-E17}) the $x$-component of the orientability
indicator for the slab space with flip $E_{17}$ takes the form
\begin{eqnarray}\label{eqdispersion-dipole-yx}
& & ^{^{(y)}}{\!} \mbox{\large $I$}_{v^2_x}^{E_{17}}({\bf x},t) = \nonumber \\
& & \hspace{.5cm} \lim_{{\bf x}^{\prime}
\to {\bf x}}\frac{p^2}{\pi^2m^2} \sum\limits_{{n_x=-\infty}}^{{\infty\;\;\prime}}
\int_0^t\int_0^t dt' dt''
\partial_y\partial_{y^{\prime}}  \frac{\Delta t^2 + r^2 -2r_x^2}{({\Delta t}^2 - r^2)^3}.
\end{eqnarray}
with $r$ defined by Eq. (\ref{separation-E17}) and  $\Delta t= t'-t''$, while  $r_x$ is
given by Eq.~(\ref{r-components-E17}). Making use of
\begin{eqnarray}\label{partial-y-yprime-yx}
& & \partial_y\partial_{y^{\prime}} \frac{\Delta t^2 + r^2 -2r_x^2}{({\Delta t}^2 - r^2)^3}
= -4(-1)^{n_x} \bigg\{ \frac{2}{({\Delta t}^2 -r^2)^3} \nonumber \\
& & \hspace{2.5cm} + 3 \frac{r^2 - r_x^2+ 6r_y^2}{({\Delta t}^2 -r^2)^4}
+ 24 \frac{(r^2 - r_x^2) r_y^2}{({\Delta t}^2 -r^2)^5} \bigg\}.
\end{eqnarray}
we find
\begin{eqnarray}\label{eqdispersion-dipole-yx-final-E17}
& & ^{^{(y)}}{\!} \mbox{\large $I$}_{v^2_x}^{E_{17}}({\bf x},t)= -\frac{4p^2}{\pi^2m^2} \sum\limits_{{n_x=-\infty}}^{{\infty\;\;\prime}}(-1)^{n_x}\bigg\{ 2I_1  \nonumber \\
& & \hspace{1.8cm} + 3 (r^2 - r_x^2 + 6r_y^2)I_2 + 24 (r^2 - r_x^2) r_y^2I_3   \bigg\},
\end{eqnarray}
where, with $\Delta t = t' - t''$,
\begin{eqnarray}\label{integral-1}
& & I_1 = {\cal I}  =  \int_0^t  \int_0^t \frac{dt^{\prime}dt^{\prime\prime}}{(\Delta t^2 -r^2)^3}
= \frac{t}{16}\bigg[ \frac{4t}{r^4 (t^2-r^2)} \nonumber \\
& & \hspace{5cm} +\frac{3}{r^5} \ln \frac{(r-t)^2}{(r+t)^2} \bigg],
\end{eqnarray}
\begin{eqnarray}\label{integral-2}
& & I_2 =  \int_0^t  \int_0^t \frac{dt^{\prime}dt^{\prime\prime}}{(\Delta t^2 -r^2)^4}
= \frac{1}{6r}\frac{\partial I_1}{\partial r} = \frac{t}{96}
\bigg[ \frac{4t(9r^2-7t^2)}{r^6 (t^2-r^2)^2} \nonumber \\
& & \hspace{5cm} - \frac{15}{r^7} \ln \frac{(r-t)^2}{(r+t)^2} \bigg],
\end{eqnarray}
\begin{eqnarray}\label{integral-3}
& & I_3 =  \int_0^t  \int_0^t \frac{dt^{\prime}dt^{\prime\prime}}{(\Delta t^2 -r^2)^5}
= \frac{1}{8r}\frac{\partial I_2}{\partial r}  \nonumber \\
& & \hspace{-.1cm} = \frac{t}{768}\bigg[
 \frac{4t(57t^4- 136r^2t^2 + 87r^4)}
{r^8 (t^2-r^2)^3} + \frac{105}{r^9} \ln \frac{(r-t)^2}{(r+t)^2} \bigg].
\end{eqnarray}
Similar  calculations lead to
\begin{eqnarray}\label{eqdispersion-dipole-yy-final-E17}
^{^{(y)}}{\!} \mbox{\large $I$}_{v^2_y}^{E_{17}}({\bf x},t) & = & -\frac{2p^2}{\pi^2m^2}
\sum\limits_{{n_x=-\infty}}^{{\infty\;\;\prime}} (-1)^{n_x}
\bigg\{ (5-3(-1)^{n_x}) I_1 \nonumber \\
& & \hspace{-1.8cm} + 6 [r^2 + (7-6(-1)^{n_x})r_y^2] I_2 + 48 [r^2 -(-1)^{n_x}r_y^2]r_y^2I_3
\bigg\}
\end{eqnarray}
and
\begin{eqnarray}\label{eqdispersion-dipole-yz-final-E17}
& & ^{^{(y)}}{\!} \mbox{\large $I$}_{v^2_z}^{E_{17}}({\bf x},t)  =  -\frac{4p^2}{\pi^2m^2} \sum\limits_{{n_x=-\infty}}^{{\infty\;\;\prime}}(-1)^{n_x}\bigg\{ 2I_1 \nonumber \\
& & \hspace{3.5cm} + 3(r^2 + 6 r_y^2 )I_2  + 24 r^2 r_y^2I_3 \bigg\}.
\end{eqnarray}
Since the coincidence limit ${\bf x}^{\prime} \to {\bf x}$ has been taken, it follows
from Eq.~(\ref{r-components-E17}) that in Eqs.~(\ref{eqdispersion-dipole-yx-final-E17}) to
(\ref{eqdispersion-dipole-yz-final-E17}) one must put
\begin{eqnarray}
\label{r-rx-ry-coincidence-E17}
 r & = & \sqrt{n_x^2a^2 + 2(1-(-1)^{n_x})y^2}, \nonumber \\
 r_x^2 & = & n_x^2a^2,
\qquad   r_y^2 = 2(1-(-1)^{n_x})y^2.
\end{eqnarray}

It can be immediately checked that, as for the point charge,  in the Minkowskian
limit ($a \to \infty$) the orientability indicator for a dipole is zero.

For the slab space $E_{16}$ the components of the dipole orientability  indicator
are obtained from those for $E_{17}$  by setting $r_x^2 = r^2, r_y=0 $,
and replacing $(-1)^{n_x}$ by $1$ everywhere. Therefore, we have
\begin{eqnarray}\label{eqdispersion-dipole-yx-final-E16}
^{^{(y)}}{\!} \mbox{\large $I$}_{v^2_x}^{E_{16}}({\bf x},t) & = & -\frac{8p^2}{\pi^2m^2}
\sum\limits_{{n_x=-\infty}}^{{\infty\;\;\prime}} I_1 ,\\
\label{eqdispersion-dipole-yy-final-E16}
^{^{(y)}}{\!} \mbox{\large $I$}_{v^2_y}^{E_{16}}({\bf x},t) & = & -\frac{4p^2}{\pi^2m^2}
\sum\limits_{{n_x=-\infty}}^{{\infty\;\;\prime}} ( I_1 + 3r^2I_2 ),\\
\label{eqdispersion-dipole-yz-final-E16}
^{^{(y)}}{\!} \mbox{\large $I$}_{v^2_z}^{E_{16}}({\bf x},t) & = & -\frac{4p^2}{\pi^2m^2}
\sum\limits_{{n_x=-\infty}}^{{\infty\;\;\prime}} (2I_1 + 3r^2I_2 ),
\end{eqnarray}
in which $r=\vert n_x \vert a$.
\begin{figure*}[tb]          
\begin{center} 
\includegraphics[width=7.1cm,height=5.9cm]{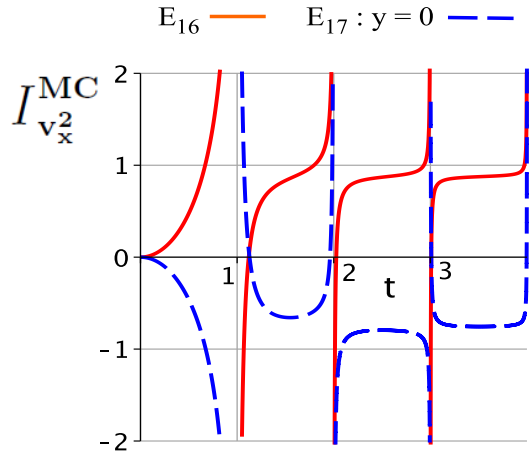} 
 \hspace{8mm}  %
\begin{minipage}[t]{\textwidth} \vspace{-1mm}
\renewcommand{\baselinestretch}{0.96}
\caption{Time evolution of the $x$-component of the  orientability
indicator in units of $p^2/m^2$
for a  point electric dipole
oriented in the flip $y$-direction  in Minkowski spacetime with
orientable $E_{16}$ and non-orientable $E_{17}$ spatial topologies, both
with compact length $a=1$.
The solid and dashed lines stand, respectively, for
the indicator  curves  
for a dipole in $3-$space with $E_{16}$ and $E_{17}$ topologies.
For the globally inhomogeneous topology $E_{17}$ the dipole is 
at $P_0= (x,0,z)$, thus
freezing out the  topological inhomogeneity. 
Both orientability indicator  curves show a periodicity, but the curve for $E_{17}$
exhibits a different kind of periodicity characterized by  a distinctive
inversion pattern.
Non-orientability is responsible for this  pattern of successive inversions,
which is absent in the indicator curve for the orientable $E_{16}$.   
\label{E16_vs_E17-fig4}  }  \vspace{2mm}
\end{minipage}
\end{center}
\end{figure*}

\vspace{5mm}
\noindent
\textbf{\small NON-ORIENTABILITY WITH A DIPOLE -- CONCLUSIONS}
\vspace{2mm}

\begin{figure*}[tb]              
\begin{center} 
\includegraphics[width=7.1cm,height=5.9cm]{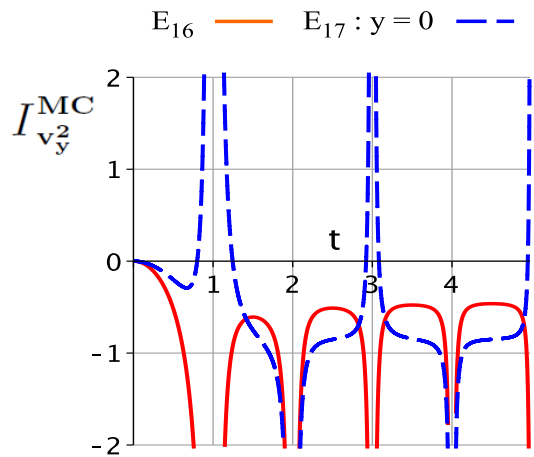} 
 \hspace{8mm}  %
\begin{minipage}[t]{\textwidth} \vspace{-1mm}
\renewcommand{\baselinestretch}{0.96}
\caption{Time evolution of the $y$-component %
of the  orientability indicator
for a point electric dipole oriented in the $y$-direction under the same
conditions as those  of Fig. \ref{E16_vs_E17-fig4}.
The orientability indicator curve for
$E_{17}$ also displays a characteristic inversion pattern but which is different
from the one for the $x$-component shown in Fig. \ref{E16_vs_E17-fig4}.
For the $y$-component of the orientability indicator the signature of non-orientability
can be recognized in the pattern of successive upward and downward
``horns'' formed by the dashed  curve.
\label{E16_vs_E17-fig5}  }  
\end{minipage}
\end{center}
\end{figure*}

\begin{figure*}[bt]                  
\begin{center} 
\includegraphics[width=7.1cm,height=5.9cm]{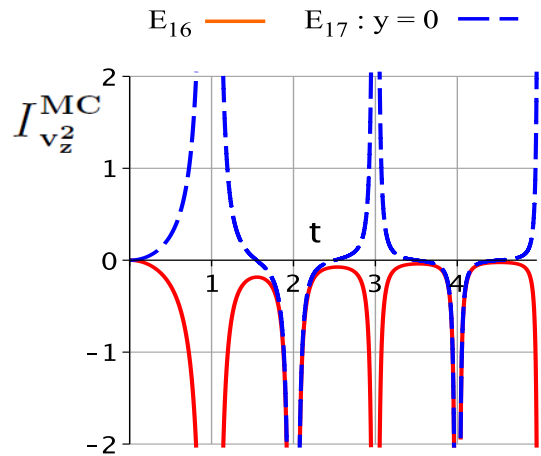}
 \hspace{8mm}  %
\begin{minipage}[t]{\textwidth} \vspace{-1mm}
\renewcommand{\baselinestretch}{0.96}
\caption{Time evolution of the $z$-component 
of the orientability indicator
for a point electric  dipole oriented in the $y$-direction under the same conditions
as those of Fig. \ref{E16_vs_E17-fig4}.
For the $z$-component of the orientability indicator the non-orientability of $E_{17}$
manifests itself by an inversion pattern similar to the one for the $y$-component
shown in Fig. \ref{E16_vs_E17-fig5}, namely the pattern of alternating upward and
downward ``horns'' formed by the dashed  curve.
\label{E16_vs_E17-fig6}  }   \vspace{-5mm} 
\end{minipage}
\end{center}
\end{figure*}

We begin by noting that the expressions (\ref{eqdispersion-dipole-yx-final-E17})%
-(\ref{eqdispersion-dipole-yz-final-E17})  and (\ref{eqdispersion-dipole-yx-final-E16})-(\ref{eqdispersion-dipole-yz-final-E16})
for 
the components of the orientability indicator for $E_{17}$ and $E_{16}$ topologies,
respectively,  are too
involved to lend themselves to a straightforward interpretation. Nevertheless,
something significant can be said: for a dipole located at $P_0 = (x,0,z)$ all components
of the orientability indicator for $E_{17}$ are different from those for $E_{16}$ because
each summand in Eqs.~(\ref{eqdispersion-dipole-yx-final-E17}),
(\ref{eqdispersion-dipole-yy-final-E17}) and (\ref{eqdispersion-dipole-yz-final-E17})
contains  the prefactor  $(-1)^{n_x}$ which is absent from the corresponding
Eqs.~(\ref{eqdispersion-dipole-yx-final-E16})~--~(\ref{eqdispersion-dipole-yz-final-E16})
for $E_{16}$.
Since not much further can be read from our equations, in order  to demonstrate our
main result, which is ultimately stated in terms of patterns of curves for the
orientability indicator, we begin by plotting figures for the components of the
orientability indicator.
Figures~\ref{E16_vs_E17-fig4} to~\ref{E16_vs_E17-fig6} come from
Eqs.~~\eqref{eqdispersion-dipole-yx-final-E17}~--~\eqref{eqdispersion-dipole-yz-final-E17}
as well as \eqref{eqdispersion-dipole-yx-final-E16}~--~\eqref{eqdispersion-dipole-yz-final-E16},
with the topological length $a=1$
and  $n_x \neq 0$ ranging from $-50$ to $50$.
In the three figures  the solid lines stand for the orientability indicator  
curves for the dipole in  Minkowski spacetime with $E_{16}$ orientable spatial topology,
whereas the dashed lines represent orientability indicator curves for the dipole located
at $P_0 = (x,0,z)$ in a $3-$space with $E_{17}$ non-orientable
topology.

In the case of the $x$-component, the time evolution curves of the orientability indicator
for $E_{16}$ and $E_{17}$, shown in Fig. \ref{E16_vs_E17-fig4}, present a common periodicity but
 with clearly distinguishable patterns. The orientability indicator curve for $E_{17}$ displays
a distinctive sort of periodicity characterized by  an inversion pattern.
Non-orientability gives rise to  this  pattern of consecutive inversions, which is not
present  for the orientable $E_{16}$.

The differences become more salient when one considers  the other two components of
the orientability indicator, shown in Figs.~\ref{E16_vs_E17-fig5} and~\ref{E16_vs_E17-fig6}.
For both of these components, the non-orientability of $E_{17}$ is disclosed by an
inversion pattern whose structure is more striking than the one for the $x$-component.
The orientability indicator curves for $E_{17}$ form a  pattern of alternating upward
and downward  
``horns'',  making  the non-orientability of $E_{17}$ unmistakably identifiable.

From the above analysis of Figs.~\ref{E16_vs_E17-fig4} to~\ref{E16_vs_E17-fig6} as
compared with the corresponding  analysis of Fig.~\ref{E16_vs_E17-fig2}, we see that
the point dipole is potentially a much more efficient non-orientability probe than the
point particle.
Furthermore, the characteristic inversion pattern exhibited by the dipole indicator curves
makes it possible to identify the non-orientability of $E_{17}$ by itself, without having
to make a comparison with the indicator curves for its orientable counterpart.  
We have  checked, although we do not show the calculations, that for a dipole located
at $P_0 = (x,0,z)$ and oriented either in the $x$-direction or in the $z$-direction,
only the $y$-component of the orientability indicator for $E_{17}$ is different
from the one for $E_{16}$. Thus, it is for a dipole oriented in the flip direction that
the non-orientability of  $E_{17}$ is most sharply exposed.

This discussion suggests  that it may be possible to  unveil
a presumed spatial non-orientability by local means, namely by the stochastic motions of
point particles caused by quantum  electromagnetic vacuum fluctuations. If the motion of
a point electric dipole is taken as probe, non-orientability  can be intrinsically discerned
by the inversion pattern of the dipole's orientability indicator curves.

To close this section, some additional words of clarification regarding the use of the
orientability indicator $\mbox{$I$}_{v^2_i}^{MC} ({\bf x},t)$ 
are fitting. We first note that, from a theoretical viewpoint,
similarly to the way we calculated the orientability indicator for $E_{17}$ and $E_{16}$
topologies
[(\ref{eqdispersion-dipole-yx-final-E17})~--~(\ref{eqdispersion-dipole-yz-final-E17}) and (\ref{eqdispersion-dipole-yx-final-E16})~--~(\ref{eqdispersion-dipole-yz-final-E16})]
one can use the  indicator's definition \eqref{new-ind} 
to derive analytic expressions and plot corresponding curves
for all eight non-orientable Euclidean topologies for a dipole.
This set of theoretical curves for the orientability indicator would form
a collection of  curve patterns that are signatures of non-orientability,
and which can be  compared with experimental curves to find out about orientability
of $3-$space.
From the experimental standpoint there are two 
main approaches to the use of  $\mbox{$I$}_{v^2_i}^{MC}$.
One is to contrast theory and observation through engineered experiments.
This would be useful to check the different expressions we have derived
for $\mbox{$I$}_{v^2_i}^{MC}$ against experimental data. 
In this case, one would need to manufacture devices that simulate non-trivial
topologies.
However, experimental apparatuses to mimic non-trivial topologies are  not
easy to prepare.%
\footnote{For a very simple nontrivial topology such as the $3-$space 
reflection orbifold a perfectly reflecting 
plane should simulate the topology.
Along these lines, one might wonder whether  many equally
spaced perfectly reflecting planes would simulate $E_{16}$. Simulation of
nontrivial topologies in laboratory is indeed a thorny issue for experimentalists, for it
depends on the physical phenomena involved. This issue is beyond the scope of the
present work, but we mention that in a different context examples of simulations
of non-trivial  topologies such as the cylinder, the torus and the
non-orientable M\"obius strip have been recently discussed~\cite{Boada-etal-2015}.}

Another experimental approach is to  use the  indicator $\mbox{$I$}_{v^2_i}^{MC}$
to directly access orientability of the spatial section of Minkowski spacetime.
First one would measure the  velocity correlation functions
$\bigl\langle \Delta v_i({\bf x},t) \Delta v_i({\bf x}^{\prime},t)\bigr\rangle^{exp}$ for
${\bf x} \neq {\bf x}^{\prime}$. Then one would subtract out the theoretically-computed correlation
functions $\bigl\langle \Delta v_i({\bf x},t) \Delta v_i({\bf x}^{\prime},t)\bigr\rangle^{SC}$
for topologically trivial (simply-connected) Minkowski spacetime, which are given in the Appendix.
The corresponding curves for this difference would be plotted in the coincidence limit
${\bf x} = {\bf x}^{\prime}$ and contrasted with each of
the eight theoretical curves mentioned in the previous paragraph.

\section{Conclusions and final remarks}  \label{Conclusion}

In general relativity and quantum field theory  spacetime is modeled as
a differentiable manifold, which is a topological space equipped with an additional
differential structure.  
Orientability is an important topological property of spacetime manifolds.
It is often assumed that the spacetime
manifold is orientable and, additionally, that it is separately time and space
orientable. The theoretical arguments usually offered to assume orientability
combine the space-and-time universality of local physical experiments%
\footnote{The space universality can be looked upon as a topological assumption
of global spatial homogeneity, which in turn rules out spatial non-orientability
of  $3-$space.}  
with physically well-defined (thermodynamically, for example) local arrow of time,
violation of charge conjugation and parity (CP violation) and CPT invariance
~\cite{Zeldovich1967,Hawking,Geroch-Horowitz-1979}.
Another theoretical
argument in favor  of a time- and
space-orientable spacetime comes from the impossibility
of having globally defined spinor fields on non-orientable
spacetime manifolds.
One can certainly use such reasonings in support of the standard assumptions
on the global structure of spacetime.
\footnote{See Ref. \cite{MarkHadley-2018}
for a dissenting point of view, and also the related Ref.~\cite{MarkHadley-2002}.}
Nevertheless, it is reasonable to expect that the ultimate answer to
questions regarding the orientability of spacetime should rely on  cosmological
observations or local experiments, or might come from a fundamental
theory of physics.

In the physics at daily and even astrophysical length and time scales, we do
not find any sign or hint of non-orientability. This being true, %
the remaining open question is whether the physically well-defined local orientations
can be extended continuously to cosmological  scales.

At the cosmological scale, one would think at first sight that to disclose  spatial
orientability one would have to make a trip around the whole $3-$space to check for
orientation-reversing paths.  
Since such a global journey across the Universe is not feasible one might
think that spatial orientability cannot be probed globally.
However, a  determination  of the spatial 
topology through the so-called circles in the sky~\cite{CSS1998}, for example,
would bring out as a bonus an answer to the $3-$space orientability problem
at the cosmological scale.
\footnote{
In the searches for these circles so far undertaken, including the
ones carried out  by the Planck Collaboration~\cite{Planck-2013-XXVI,Planck-2015-XVIII},
no statistically significant pairs of matching circles have been found (see
Ref.~\cite{Vaudrevange-etal-12} for the most extensive search yet, and
also references therein for the other searches).
These negative observational results, however,  are not sufficient to exclude the
possibility that the Universe has a detectable (orientable or non-orientable)
nontrivial topology (see Ref.~\cite{Gomero-Mota-Reboucas-2016} for some limits of
these searches).  }

There is no reason in the classical physics of point-like  particles to
force a spacetime manifold $\mathcal{M}_4 = \mathbb{R}\times M_3$ to be
orientable. However, the situation  changes once fermions, for example,
are considered at the  quantum level.
Indeed, in quantum field theory one uses spinors to describe
fermions. This fact inevitably leads to the inclusion
of spinors as desirable objects (\textsl{locally}) in the spacetime
$\mathcal{M}_4$ on physical grounds.
However, the \textsl{global} question, which ultimately is not a physical requirement,
is left open and the assumption of existence of 
\textsl{globally} defined spinor fields, which rules out non-orientable
spacetimes~\cite{Penrose-1968,Penrose-Hindler1986,Geroch-1968,Geroch-1970},
comes from theoretical arguments that combine this local physics
requisite with the space-and-time universality of the basic local rules
of physics.  
For spacetimes  of the form $\mathcal{M}_4 = \mathbb{R}\times M_3$,
the spatial universality (see related comment on p.288 of
Visser~\cite{Visser-1996}), can be mathematically translated in terms
of global (topological) homogeneity of $M_3$, which
in turn rules out spatial non-orientability of the  $3-$space
since there is no globally homogeneous non-orientable quotient
manifold $M_3$.
As a matter of fact, the spatial universality alone as captured in
topological terms rules out not only the non-orientable but
any topologically inhomogeneous spatial sections $M_3$.
In this way, in arguing for the universality of the local physics, which
would lead to the existence of globally defined  spinor fields, 
we are actually making a topological assumption of global homogeneity of
$3-$space, which itself excludes the $8$ non-orientable $3-$manifolds 
and also the remaining $6$ other globally inhomogeneous and orientable
spatial sections.%
\footnote{In technical terms, the ultimate mathematical
point that leads to the fact that  we cannot have globally defined
spinor fields in non-orientable manifolds lies in bringing together the
two-valuedness of spinors with the flip holonomy. Actually, this mismatching
occurs even in some orientable, but inhomogeneous, quotient manifolds $M_3$,
whose covering or holonomy group contains at least one $\gamma \in \Gamma$
which is a proper screw motion $R(\alpha,\mathbf{\widehat{u}})$ with
$\alpha \neq 0$ (cf. Section~II).}
In brief, one can only have global spinor fields in spacetimes
$\mathcal{M}_4 = \mathbb{R}\times M_3$ whose spatial section is globally
(topologically) homogeneous. This obviously excludes the topologically
inhomogeneous $M_3$, whose set contains all non-orientable $3-$space
manifolds.

In this paper we have addressed the question as to whether
electromagnetic quantum vacuum  fluctuations 
can be used to bring out  the spatial orientability of
Minkowski spacetime.
To this end, we have studied the stochastic motions of point-like particles
under quantum electromagnetic fluctuations in Minkowski spacetime with the
orientable slab space ($E_{16}$) and the non-orientable slab space with flip
($E_{17}$) topologies (cf. Tables~\ref{Tb-4-Orient_and_Non_orient}
and~\ref{Tb-Spatial-separation}).

For a point charged particle, we have derived analytic expressions
for the orientability indicator, namely
Eqs.~\eqref{dispersion-x-E17}~--~\eqref{r-ri-coincidence-r-bar-z}
for
$E_{17}$ space topology, and
Eqs.~\eqref{dispersion-orientable-x-E16}~--~\eqref{dispersion-orientable-yz-E16}
for  $E_{16}$ space topology. From these equations we have made
Fig.~\ref{E16_vs_E17-fig1} and Fig.~\ref{E16_vs_E17-fig2}.
Using these equations and figures we have shown that it is possible to distinguish
the orientable from the non-orientable topology by contrasting the time
evolution of the respective orientability indicators  along the  flip direction of $E_{17}$.

In spite of being a significant result in that it makes apparent
the power of our approach to access orientability through electromagnetic 
quantum vacuum fluctuations, it is desirable, however, to be able to decide about
the orientability of a given spatial manifold in  itself.
To tackle this question, motivated by a dipole's directional properties,
we have then examined  whether the study of stochastic motions of a point-like electric
dipole would be more effective for testing the non-orientability of a generic
$3-$space individually, i.e. without having to make a comparison of the results
for an orientable space with those for its non-orientable counterpart. 

To this end, we have derived the expressions for the orientability indicators given by
equations~\eqref{eqdispersion-dipole-yx-final-E17}~--~\eqref{eqdispersion-dipole-yz-final-E17} for
the dipole oriented in the flip direction in the non-orientable  $3-$space with $E_{17}$ topology,
and equations~\eqref{eqdispersion-dipole-yx-final-E16}~--~\eqref{eqdispersion-dipole-yz-final-E16}
for the dipole in the orientable $3-$space with $E_{16}$ topology. From these equations we have
calculated and plotted Figures~\ref{E16_vs_E17-fig4} to \ref{E16_vs_E17-fig6}.
As a result of the  analysis of these equations and figures we have
found that there exists a characteristic inversion pattern exhibited by the
 orientability indicator curves in the case of   
$E_{17}$, signaling that the non-orientability of $E_{17}$
might be identified per se. The inversion pattern of the  orientability indicator curves
for the dipole is a signature of  the reflection holonomy, and ought to be
present in the orientability indicator  curves for the dipole in all remaining seven
non-orientable 
topologies with flip, namely  the four Klein spaces ($E_{7}$ to $E_{10}$) and those
in the chimney-with-flip class ($E_{13}$ to $E_{15}$).%
\footnote{See Refs.~\cite{Riazuelo-et-el03,Fujii-Yoshii-2011}
for the symbols, names and properties of Klein spaces and chimney-with-flip
families of non-orientable topologies.}
Clearly the inversion patterns for the electric dipole change with the
associated topology:  different topologies give rise to
curves for the orientability indicator with distinct inversion patterns.

A short  comment on how  our
results extend to the quantum level is fitting. We have tacitly
assumed that the charged object of interest is described by a
well-localized wave packet. Our classical analysis is justified
because it has been argued [16] that under suitable conditions the
dispersion due to wave packet spreading is negligible as compared
with the dispersion brought about by vacuum fluctuations
of the electromagnetic field.

Observation of physical phenomena and experiments are fundamental to our
understanding of the physical world.
Our results indicate  that it is potentially possible to locally  unveil
spatial non-orientability through  stochastic motions of point-like particles
under electromagnetic quantum vacuum fluctuations.   
In this way, the present paper may be looked upon as an intimation of a conceivable
way to locally probe the spatial orientability of Minkowski empty spacetime.
A different statistical indicator of the stochastic motions can certainly
be conceived, but perhaps the most important point raised in this work  is
that the spatial orientability of spacetime can be locally probed 
by inquiring  quantum vacuum electromagnetic fluctuations about the motion
of point-like  ``charged'' particles.

\begin{acknowledgements}
M.J. Rebou\c{c}as acknowledges the support of FAPERJ under a CNE E-26/202.864/2017 grant,
and thanks CNPq for the grant under which this work was carried out.
We also thank  C.H.G.  Bessa for fruitful discussions and for his help with the figures.
M.J.R. is also grateful to A.F.F. Teixeira for interesting comments, and also for reading
the manuscript and indicating typos.
\end{acknowledgements}


\appendix
\label{velocity-correlation}
\section{Velocity correlation functions for the simply-connected $\mathbf{\,3-}$space} 

In this appendix we give the velocity correlations functions for simply-connected
spatial section of Minkowski spacetime that are referred to on the last paragraph
of Section \ref{Syst-montion},
for both the point particle and the point dipole. These are calculated as follows.
First replace  $D$ in equation (\ref{eqdif-0}) by $D_0$ as given by (\ref{eqren}),
thus obtaining the electric field correlation functions for the topologically trivial
(simply connected) Minkowski spacetime. Then insert these electric field correlation
functions with ${\bf x} \neq {\bf x}^{\prime}$ into the right-hand side of
equation~(\ref{eqdispersion1}), for the particle, and  equation~(\ref{eqdispersion-dipole}),
for the dipole.  With
\begin{equation}
\label{rx-ry-rz-ap}
r_x = x - x^{\prime}\,, \quad r_y = y - y^{\prime} , \quad r_z = z - z^{\prime}\,,
\end{equation}
and
\begin{equation}\label{r-ri-coincidence-r-ap}
r = \sqrt{( x - x^{\prime})^2 + ( y - y^{\prime})^2 + ( z - z^{\prime})^2},
\end{equation}
we have, for the particle:
\begin{eqnarray}\label{correlation-x-E17-ap}
 \Bigl\langle \Delta v_x({\bf x},t) \Delta v_x({\bf x}^{\prime},t)\Bigr\rangle^{SC} & = &
 \frac{q^2 t}{8\pi^2 m^2 r^5 (t^2 - r^2)}
\bigg\{ 4r(r_y^2+r_z^2)t \nonumber \\
& & \hspace{-.9cm} + (r^2-3r_x^2)(t^2-r^2) \ln \frac{(r-t)^2}{(r+t)^2} \bigg\}, \\
\label{correlation-y-E17-ap}
\Bigl\langle \Delta v_y({\bf x},t) \Delta v_y({\bf x}^{\prime},t)\Bigr\rangle^{SC} &  =  &
 \frac{q^2 t}{8\pi^2 m^2 r^5 (t^2 - r^2)}
\bigg\{ 4r(r_x^2+r_z^2)t  \nonumber \\
& & \hspace{-.9cm} + (r^2 -3r_y^2)(t^2-r^2) \ln \frac{(r-t)^2}{(r+t)^2} \bigg\}, \\
\label{correlation-z-E17-ap}
\Bigl\langle \Delta v_z({\bf x},t) \Delta v_z({\bf x}^{\prime},t)\Bigr\rangle^{SC} &  =  &
  \frac{q^2 t}{8\pi^2 m^2 r^5 (t^2 - r^2)}
\bigg\{ 4r(r_x^2+r_y^2)t  \nonumber \\
& & \hspace{-.9cm} + (r^2 -3r_z^2)(t^2-r^2) \ln \frac{(r-t)^2}{(r+t)^2} \bigg\}.
\end{eqnarray}
The corresponding results for the dipole oriented in the $y$-direction are
\begin{eqnarray}\label{correlation-dipole-yx-final-E17-ap}
 ^{^{(y)}}{\!} \Bigl\langle \Delta v_x({\bf x},t)
\Delta v_x({\bf x}^{\prime},t)\Bigr\rangle^{SC} & = & -\frac{4p^2}{\pi^2m^2} \bigl[ 2I_1
\nonumber \\
& & \hspace{-2.1cm} + 3 (r^2 - r_x^2 + 6r_y^2)I_2
  + 24 (r^2 - r_x^2) r_y^2I_3 \bigr], \\
\label{correlation-dipole-yy-final-E17-ap}
^{^{(y)}}{\!} \Bigl\langle \Delta v_y({\bf x},t) \Delta v_y({\bf x}^{\prime},t)
\Bigr\rangle^{SC} & = & -\frac{4p^2}{\pi^2m^2}
\bigl[  I_1 + 3 (r^2 + r_y^2) I_2 \nonumber \\
& & \hspace{0.8cm}  + 24 (r^2 -r_y^2)r_y^2I_3 \bigr],\\
\label{correlation-dipole-yz-final-E17-ap}
^{^{(y)}}{\!} \Bigl\langle \Delta v_z({\bf x},t) \Delta v_z({\bf x}^{\prime},t)
\Bigr\rangle^{SC} & = & -\frac{4p^2}{\pi^2m^2} \bigl[ 2I_1  \nonumber \\
& & \hspace{-2cm} + 3(r^2 - r_z^2 + 6 r_y^2 )I_2
 + 24 (r^2 -r_z^2) r_y^2I_3 \bigl],
\end{eqnarray}
where the integrals $I_1$, $I_2$, $I_3$ are given by equations~(\ref{integral-1})
to~(\ref{integral-3}).



\end{document}